\documentclass[prl,aps,showpacs,twocolumn,unsortedaddress]{revtex4-1}
\usepackage{graphicx,bm}
\usepackage{amssymb}
\usepackage{epsfig}
\usepackage{epsf}
\usepackage[usenames]{color}

\begin{document}

\title{Anomalous Hall conductivity from the dipole mode of
spin-orbit-coupled cold-atom systems}

\author{E. van der Bijl}
\author{R.A. Duine}
\affiliation{Institute for Theoretical Physics, Utrecht
University, Leuvenlaan 4, 3584 CE Utrecht, The Netherlands}

\date{\today}

\begin{abstract}
Motivated by recent experiments [Lin {\it et al.}, Nature {\bf 417}, 83 (2011)] that engineered spin-orbit coupling in ultra-cold mixtures of bosonic atoms, we study the dipole oscillation of trapped spin-orbit-coupled non-condensed Bose and Fermi gases. We find that different directions of oscillation are coupled by the spin-orbit interactions. The phase difference between oscillatory motion in orthogonal directions and the trapping frequencies of the modes are shown to be related to the anomalous Hall conductivity. Our results can be used to experimentally determine the anomalous Hall conductivity for cold-atom systems.
\end{abstract}


\maketitle

% definitions
\def\bx{{\bm x}}
\def\bk{{\bm k}}
\def\bK{{\bm K}}
\def\bq{{\bm q}}
\def\br{{\bm r}}
\def\bp{{\bm p}}
\def\bM{{\bm M}}
\def\bs{{\bm s}}
\def\bB{{\bm B}}
\def\bj{{\bm j}}
\def\bF{{\bm F}}
\def\id{{\rm d}}

\def\br{{\bm r}}
\def\bv{{\bm v}}

\def\half{\frac{1}{2}}
\def\args{(\bm, t)}

{\it Introduction.} --- Transport phenomena play a crucial role in understanding and characterizing condensed-matter systems. Two of these phenomena, the Hall effect and the anomalous Hall effect (AHE) were both discovered in the late 19$ ^{\rm th}$ century. That the Hall effect is due to the Lorentz force has been understood since those days. The AHE, a transverse voltage or current present in ferromagnets in the absence of a magnetic field, is related to spin-orbit (SO) coupling and has proven much more challenging to understand (for a review see Ref. \cite{Nagaosa2010}).  Since SO coupling is responsible for the AHE, anomalous-Hall-like effects should also be present for particles that do not carry charge, and, indeed, such effects are observed for magnons~\cite{Onose2010}, phonons~\cite{Strohm2005,Sheng2006}, and photons~\cite{Onoda2004}. Although these effects were observed using heat currents, and they should thus be called anomalous Righi-Leduc effects, their physical mechanism is similar to that of the AHE.

In this Letter we consider the AHE in homogeneous and harmonically-trapped cold-atom systems. (The AHE was considered in cold-atom systems in the presence of an optical lattice in two dimensions by Dudarev~\textit{et al.}~\cite{Dudarev2004}.) As the atoms are neutral, the AHE here refers to a mass current perpendicular to an applied force in the absence of a Coriolis force. (For cold-atom systems rotation and the resulting Coriolis force play the role of a magnetic field and the Lorentz force.) Our investigation is motivated by the recent experiment by Lin \textit{et al.}~\cite{Lin2011} who engineered spin-orbit coupling in a Bose-Einstein condensate \cite{Galitski2008,Zhai2010} with lasers. This experiment is one of the latest achievements in studying phenomena known from solid-state physics in a cold-atom setting. Other examples are the Mott-insulator-to-superfluid phase transition \cite{Greiner2002}, Bardeen-Cooper-Schrieffer superfluidity \cite{Jin2003}, the Berezinskii-Kosterlitz-Thouless phase transition \cite{Dalibard2006,Cornell2007}, and Anderson localization \cite{Aspect2008}. Important features of cold-atom systems are that new regimes of physics (as compared to solid-state systems) can be explored, and their great amount of tunability. Furthermore, cold-atom systems are in principle disorder free and have a well-known microscopic description making it worthwhile to undertake a detailed comparison between theory and experiment, whereas in solid-state materials typically a multitude of effects play a role which makes modeling harder.

In the case of the AHE, for example, the difficulty in understanding the effect lies in part in the interplay between so-called \textit{intrinsic} and \textit{extrinsic} contributions. Intrinsic contributions come from spin-orbit coupling effects in the bandstructure, whereas extrinsic contributions arise from disorder. A recent theoretical advancement in the understanding of the AHE is the semi-classical description in terms of equations of motion for Bloch wavepackets~\cite{Chang1995,*Jungwirth2002, Sinitsyn2008}. In this description the intrinsic contribution to the AHE stems from anomalous-velocity contributions to these semi-classical equations of motion~\cite{Karplus1954}. In modern language, this anomalous velocity results from the Berry-phase curvature of the Bloch bands that in turn is determined by the topology of the band structure. The relation between bandstructure topology and the Hall conductivity was first emphasized by Thouless \textit{et al.}~\cite{Thouless1982}, and has regained interest with the very recent discovery of topological insulators \cite{KaneMele2005,*Roth2009,*Hasan2010}.

In a typical cold-atom experiment steady-state currents are not readily created and transport coefficients can be measured only indirectly. In this Letter we show that the anomalous Hall conductivity can be obtained from the properties of the dipole oscillation of a cloud of spin-orbit coupled cold atoms that is trapped in an external harmonic trapping potential. The dipole mode is a collective oscillation of the center-of-mass of the cloud. According to Kohn's theorem \cite{Kohn1961}, the frequencies of the dipole oscillation are equal to the trap frequencies. The SO coupling, however, breaks the harmonic nature of the system and as a result Kohn's theorem for the dipole modes does not hold. We find that spin-orbit coupling modifies the oscillation frequencies and that different directions of oscillation are coupled by the spin-orbit interactions. The phase difference between oscillatory motion in different directions and the mode frequencies turn out to be related to the anomalous Hall conductivity. This result can be used to experimentally determine the anomalous Hall conductivity for cold-atom systems. Below we detail the semi-classical Boltzmann approach on which our findings are based, determine the anomalous Hall conductivity for homogeneous non-condensed Bose and Fermi gases, and show how this conductivity can be obtained from the dipole oscillation of trapped atomic gases.

{\it Semi-classical equations of motion} --- We consider spin-$1/2$ atoms with mass $m$ trapped in an external potential $V^{ex}(\bx)$ in the presence of a generic spin-orbit coupling. The Hamiltonian is
\begin{equation}\label{eq:HamSOC}
\hat{H} = \frac{\hat{\bp}^2}{2m}+V^{ex}(\hat{\bx})-\bM(\hat{\bp})\cdot{\bm \tau}~,
\end{equation}
with $\hat{\bp}$ and $\hat{\bx}$ the momentum and position operators of the particles and ${\bm \tau}$ the vector of Pauli matrices. The last term describes the SO coupling, that for spin one-half particles is without loss of generality given in terms of a momentum-dependent effective magnetic field $\bM$.

At the semi-classical level we consider in first instance the dynamics of the expectation values of the position $\bx=\langle\hat{\bx}\rangle$, momentum, $\bp=\langle\hat{\bp}\rangle$, and spin  $\bs=\hbar\langle {\bm \tau} \rangle/2$ degrees of freedom. We obtain the Heisenberg equations of motion
\begin{eqnarray}
\dot{\bx} &=&  \frac{\bp}{m}-\frac{2}{\hbar}\frac{\partial \bM}{\partial \bp}\cdot \bs~;\label{eq:dxdt}\\
\dot{\bp} &=&  -\frac{\partial V^{ex}}{\partial \bx}~;\label{eq:dpdt}\\
\dot{\bs} &=& \bs \times\frac{\bM}{\hbar} ~. \label{eq:dsdt}
\end{eqnarray}
We proceed by assuming that the spin degree of freedom is much faster than the motion of the particles. Thus we let the spin follow the effective magnetic field $\bM$ adiabatically, and only allow for a small misalignment between the spin and the effective magnetic field that is first order in time-derivatives of the orbital dynamics. This approach is essentially exact in the linear-response regime. Hence, we solve the equation for the spin degree of freedom Eq.~(\ref{eq:dsdt}) up to first order in time-derivatives by $\bs\propto\sum_i {\bm m}+\frac{\hbar}{|\bM|}({\bm m}\times\frac{\partial {\bm m}}{\partial \bp_i})\cdot\dot{\bp}_i$, with ${\bm m}(\bp(t))$ the unit vector in the direction of $\bM$. For spins opposite to the field the result is $-\bs$. Insertion of the result for $\bs$ into Eq.~(\ref{eq:dxdt}) gives~\cite{Chang1995,*Jungwirth2002, Sinitsyn2008}
\begin{equation}
\dot{\bx}_{k} = \frac{\partial \epsilon_{\bp,k}}{\partial \bp}+k\dot{\bp}\times\bB(\bp)~,\label{eq:dxdtAN}
\end{equation}
where the band index $k$ distinguishes between atoms with spin parallel ($+$) or antiparallel ($-$) to the field $\bM$. Furthermore, the dispersion is given by $\epsilon_{\bp,k}=\bp^2/2m -k |\bM(\bp)|$, and the vector field $\bB_c(\bp)=\hbar\sum_{a,b\in\{x,y,z\}}\epsilon^{abc}\left(\partial {\bm m}/\partial \bp_a\times \partial {\bm m}/\partial \bp_b\right)\cdot {\bm m}/2$ determines the anomalous velocity contribution. We refer to $\bB$ as the Berry magnetic field. 

{\it Boltzmann equation and anomalous Hall conductivity.} ---We proceed by calculating the anomalous Hall conductivity for a homogeneous gas from the Boltzmann equation for the distribution function $f_{k}(\bx,\bp,t)$ for atoms in band $k$, that is given by
\begin{equation}\label{eq:Boltzmann}
\frac{\partial f_{k}}{\partial t}+\dot{\bx}_{k}\cdot \frac{\partial f_{k}}{\partial\bx}+\dot{\bp}\cdot \frac{\partial f_{k}}{\partial\bp}=0~,
\end{equation}
where we ignored collisions as the intrinsic anomalous Hall conductivity does not depend on relaxation \cite{Nagaosa2010}. In the above, $\dot{\bx}_{k}$ and $\dot{\bp}$ are given by Eq. (\ref{eq:dxdtAN}) and Eq. (\ref{eq:dpdt}) respectively. We consider a steady-state situation with a constant applied force $\bF=-\partial V^{\rm ex}/\partial \bx$ acting equally on atoms in both bands, and define the conductivity tensor ${\bm \sigma}$ by $\bj={\bm \sigma} \cdot \bF$, where $\bj$ is the particle current density which is given by $\bj \equiv \sum_k \int \frac{\id^3\bp}{(2\pi \hbar)^3}f_{k}(\bp)\dot{\bx}_{k}$. The anomalous Hall conductivity $\sigma_{\rm AH}$ is the off-diagonal component of this conductivity tensor. The solution of the Boltzmann equation leads to the anomalous Hall conductivity
\begin{equation}\label{eq:sigmaIAH}
\sigma_{{\rm AH}} =\sum_{k\in\{+,-\}}\int \frac{\id^3\bp}{(2\pi \hbar)^3} k N(\epsilon_{\bp,k}) B_z(\bp)~,
\end{equation}
where $N(\epsilon)=[e^{(\epsilon-\mu)/k_BT}\pm 1]^{-1}$  with $k_{\rm B}T$ the thermal energy and $\mu$ the chemical potential, is the Fermi-Dirac ($+$) or Bose-Einstein ($-$) distribution function that applies for fermions or bosons, respectively. The above expression for the anomalous Hall conductivity is the intrinsic contribution due to SO coupling effects in the band structure. In cold-atom atom systems there is, unless engineered \cite{Aspect2008}, no disorder and thus extrinsic contributions are absent.

So far we have considered a generic SO coupling. In order to make a connection with experiments we will now consider a Rashba-Dresselhaus \cite{Rashba1960,*Dresselhaus1955} form of the SO coupling so that the effective magnetic field reads
\begin{equation}
\bM(\bp) = \left(\frac{\alpha}{\hbar}\bp_y-\frac{\beta}{\hbar}\bp_x,-\frac{\alpha}{\hbar}\bp_x+\frac{\beta}{\hbar}\bp_y,\frac{\Delta}{2}\right)^T
\end{equation}
where $\alpha$ and $\beta$ are the coupling constants for Rashha and Dresselhaus SO-coupling respectively, and $\Delta$ is a spin-splitting energy. We then find for the Berry magnetic field
\begin{equation}\label{eq:Bp}
\bB(\bp) = \frac{4(\alpha^2-\beta^2)\Delta\hbar^2}{(\Delta^2\hbar^2 +4 (\alpha^2 + \beta^2)\bp^2-16 \alpha \beta \bp_x \bp_y )^{3/2}}  {\bm \hat{z}}~.
\end{equation}
Note that in the experiments by Lin~\textit{et.al.} \cite{Lin2011} an equal amount of Rashba and Dresselhaus coupling was realized, i.e., $\alpha=\pm\beta$. It follows that in this specific case $\bB(\bp)=0$ and that $\sigma_{AH}=0$~\cite{Schliemann2003}. It is however experimentally straightforward to consider a more general SO coupling \cite{Lin2011}. The anomalous Hall conductivity vanishes in the absence of a spin splitting $\Delta$, in agreement with the fact that the AHE occurs in ferromagnets.

\begin{figure}[!t]
\begin{center}
\includegraphics[width=1.00\linewidth]{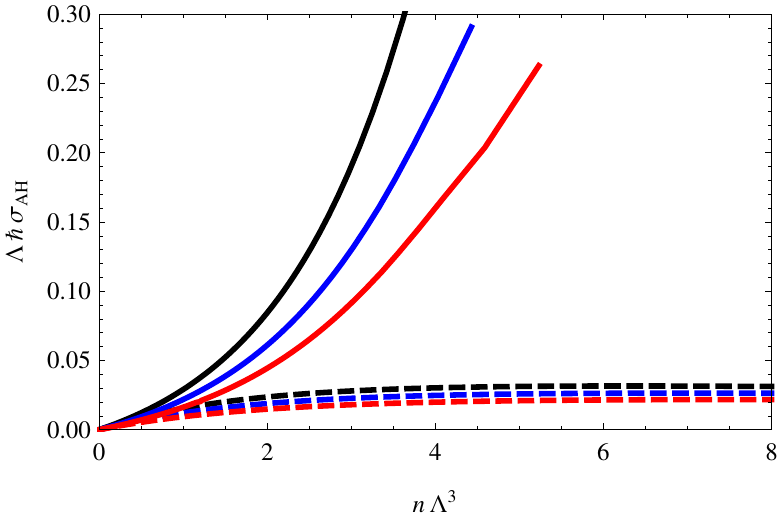}
\caption{(Color online) The anomalous Hall conductivity for bosons (solid) and fermions (dashed) as a function of $n\Lambda^3$. The lines correspond from top to bottom to $\alpha/\beta=( \infty,2,1.5 )$ where $\alpha^2+\beta^2=m\Lambda / \hbar^2 $. The spin splitting energy $\Delta=0.2 k_{\rm B}T$.}  \label{fig:sigmaIAHBosons}
\end{center}
\end{figure}

In Fig. \ref{fig:sigmaIAHBosons} we show results for the anomalous Hall conductivity of bosons and fermions. We only show the results for $\alpha>\beta$ since $\sigma_{\rm AH}(\pm \alpha,\pm \beta)=\sigma_{\rm AH}(\alpha,\beta)$ and $\sigma_{\rm AH}(\alpha,\beta)=-\sigma_{\rm AH}(\beta,\alpha)$. The results shown for bosons are above the critical temperature for Bose-Einstein condensation (this temperature depends on $\alpha, \beta$ and $\Delta$). The anomalous Hall conductivity is independent of temperature in the degenerate ($n\Lambda^3\gg1$) limit for fermions, with $n$ the density and $\Lambda\equiv(2\pi\hbar^2/mk_{\rm B}T)^{1/2}$ the de Broglie wavelength, as expected.

{\it Collective Modes.}--- We now study the dipole oscillation of an atomic cloud of $N_a$ atoms in
an anisotropic harmonic trapping potential of the form $V^{{\rm ex}}(\bx) = \frac{m}{2} \left(\omega_r^2(x^2+y^2)+\omega_z^2 z^2\right)$, with $\omega_r$ and $\omega_z$ the trapping frequencies. The dipole ($l=1$) oscillations are pure translations of the cloud with no changes in its internal structure that are described by the equations of motion for the center of mass position $\bx_0\equiv\frac{1}{N_a}\sum_k\int \id \bx \int \id \bp/(2\pi\hbar)^3  f_{k} \bx_{k}$ and velocity $\bv_0\equiv\frac{1}{N_a}\sum_k\int \id \bx \int \id \bp/(2\pi\hbar)^3 f_{k} \dot{\bx}_{k}$. Hence, we make the following \textit{ansatz} for the distibution function, $f_{k}(\bx,\bp,t)=n_k(\bx-\bx_0(t),\bp-m\bv_0(t))$, where $n_{k}(\bx,\bp)=N(\epsilon_{\bp,k}-\mu(\bx))$ is the Bose-Einstein or Fermi-Dirac distribution function in the local-density approximation, with $\mu(\bx)=\mu-V^{\rm ex} (\bx)$. From the Boltzmann equation we obtain the equations of motion for the center-of-mass coordinates. For small oscillations, we linearize these equations of motion resulting in
\begin{eqnarray}
\dot{\bx}_0 &=& \bar{H}\cdot \bv_0- \nabla V^{ex}(\bx_0)\times\bar{\bB};\\
m\dot{\bv}_0 &=& -\nabla V^{ex}(\bx_0) \label{eq:comv}
\end{eqnarray}
where $\bar{H}$ is proportional to the Hessian matrix of the dispersion and $\bar{\bB}$ is the Berry magnetic field averaged over the trap, which are given by
\begin{eqnarray*}
\bar{\bB} &=&\frac{1}{N_a} \sum_{k\in\{+,-\}} \int \id^3\bx \int\frac{\id^ 3\bp}{(2\pi \hbar)^3} k n_{k}(\bx,\bp) \bB(\bp);\\
\bar{H}_{ij} &=&\frac{1}{N_a} \sum_{k\in\{+,-\}} \int \id^3\bx \int\frac{\id^3\bp}{(2\pi\hbar)^3} n_{k}(\bx,\bp) H_{ij,k}(\bp),
\end{eqnarray*}
with  $H_{ij,k}\equiv m \partial^2 \epsilon_{\bp,k}/\partial\bp_i\partial\bp_j$. For the mode in the $z$-direction we obtain the result $\omega=\omega_z$ as predicted by Kohn's theorem and expected since the SO coupling only affects the dynamics in the $x-y$ plane. The modes in this plane have frequencies given by
\begin{equation}\label{eq:modes}
\omega_{\pm} =  \omega_r\sqrt{A \pm \sqrt{(\bar{H}_{xy}^2-\bar{H}_{xx}\bar{H}_{yy})+A^2}},
\end{equation}
with $A= (\bar{H}_{xx}+\bar{H}_{yy}+\bar{\bB}_z^2 m^2\omega_r^2)/2$.
\begin{figure}[!t]
\begin{center}
\includegraphics[width=1.00\linewidth]{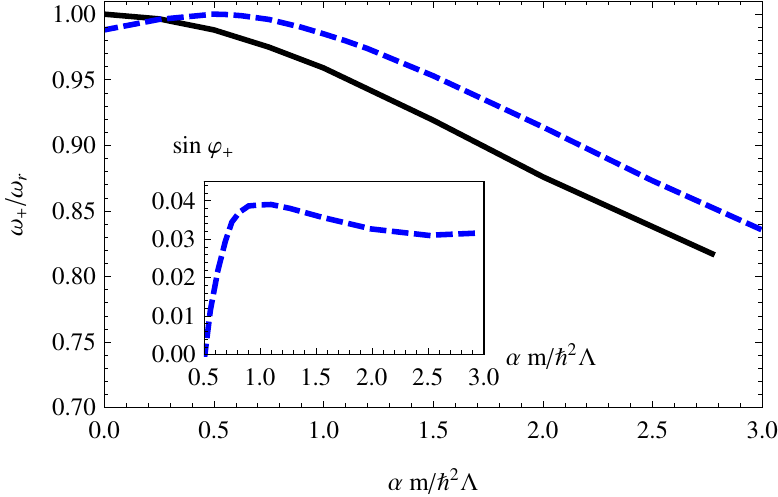}
\caption{(Color online) Dipole mode frequencies as given by Eq. (\ref{eq:modes}). The solid line is calculated for $\beta=0$, the dashed lines correspond to $\beta =0.5 m/\hbar^2\Lambda$. The spin splitting is $\Delta=0.2 k_{\rm B}T$. The number of particles $N_a=1.8\times10^5$ and temperature $T=200$ nK. The inset shows the $\sin \varphi$ as a function of $\alpha$} \label{fig:omegaphi}
\end{center}
\end{figure}
When there is no SO coupling, $\alpha, \beta=0$ we find $\omega=\omega_r$ as predicted by Kohn's theorem. For nonzero SO coupling the double degeneracy of this mode is lifted. The eigenmodes of oscillation are given by $\bx^{\pm}_0(t) = \left(x_1 \sin(\omega_{\pm} t+\varphi_{\pm}) ,x_2 \sin(\omega_{\pm} t)\right)^T$ with $\sin\varphi_{\pm}=m\bar{\bB}_z\omega_{\pm}/(\bar{H}_{xy}^2+m^2\bar{\bB}_z^2\omega_{\pm}^2)^{1/2}$. In Fig. \ref{fig:omegaphi} we show the mode frequency $\omega_{+}$ and angle $\varphi_{+}$ as a function of $\alpha$, in the special case $\alpha=\pm\beta$ we have $\bar{\bB}_z=0$ and find $\varphi_{\pm}=0$ as shown in the inset. Another special case occurs when the SO coupling is of the pure Rashba or Dresselhaus form. Then $\bar{H}_{xy}$ vanishes which results in $\varphi=\pi/2$. The phase difference between the two different directions of oscillations is determined by the Berry magnetic field. We can relate the average of the Berry magnetic field over the trap to the anomalous Hall conductivity for a homogeneous gas with a density equal to the central density $n_0$ of the trapped cloud $\bar{\bB}_z \simeq \sigma_{{\rm AH}}(n_0)/n_0$, where $\sigma_{\rm AH}(n_0)$ is given by Eq.~(\ref{eq:sigmaIAH}). This conductivity can therefore be experimentally determined by measuring the frequencies $\omega_{\pm}$ or the phase differences $\varphi_{\pm}$ of the modes.

{\it Discussion and conclusions} --- We have studied the dipole oscillation of a trapped gas of spin-orbit coupled cold-atoms, and found that these oscillations can be used as an experimental probe for the anomalous Hall effect. In the experiments by Lin~\textit{et.al.}\cite{Lin2011} the SO coupling strength is $\alpha,\beta\simeq  m \Lambda/\hbar^2$ and the Zeeman spin-splitting is $\Delta\simeq 0.2 k_B T$, using a temperature of $T \simeq 200$ nK. Taking values for $\alpha$ and $\beta$ of this order, we find that the angle $\varphi_{+} \simeq0.03$ and that $(\omega_{\pm}-\omega_r)/\omega_r \simeq 10\%$ which appear to be observable. 

Up to this point we have not considered collisions between the atoms leading to damping of collective oscillations. The harmonic nature of our system is explicitly broken by the SO coupling leading to relaxation of the center-of-mass motion of the cloud (such relaxation is absent when the SO coupling is zero and Kohn's theorem prevails). This can be described phenomenologically by adding a term $-\bv_0/\tau$ on the right-hand-side of Eq. (\ref{eq:comv}) which would lead to damping of the dipole modes but does not affect the anomalous Hall conductivity. The frequencies of the damped system are $\Omega_{\pm} = \omega_{\pm} + i \gamma_{\pm}$ with the damping rate $\gamma_{\pm}$, up to first order in $1/\tau$, given by
\[
\gamma_{\pm} = \frac{1}{\tau}\left(\frac{\omega_{\pm}^2+\bar{H}_{xx}\omega_r^2}{2\omega_{\pm}^2+2\bar{H}_{xx}\omega_r^2+\bar{\bB}_z^2m^2\omega_r^4}-1\right).
\]
We note that the relaxation time $\tau$ can in principle be calculated from the Boltzmann equation but considering this, given the above remarks regarding its importance, is beyond the scope of the present paper.

In the adiabatic approximation that leads to the semi-classical equations of motion, spin directions transverse to the magnetic field $\bM (\bp)$ are taken into account approximately as they give rise to the anomalous velocity terms. One could go beyond this adiabatic approximation and consider the $(2\times2)$-distribution function $f_{\sigma \sigma'}(\bp)$ that allows for all possible spin directions. We have checked, by solving the Boltzmann equation for this distribution function in the collisionless limit, that our results for the anomalous Hall conductivity and the phases $\varphi_{\pm}$ are not altered.

Possible extensions of this work are to consider the partially Bose-Einstein condensed phase for bosons, and the situation without Zeeman spin splitting $\Delta$. In the latter case the AHE is absent, but there will be a spin Hall effect \cite{Murakami2003, Sinova2004} that can be probed via the spin-dipole mode. (The spin Hall effect for cold atoms was proposed by Zhu~\textit{et al.} \cite{Zhu2006} for a cloud falling due to gravitation.) We also intend to investigate the effects of spin-orbit coupling on other collective modes, in particular the quadrupole oscillation. 

\acknowledgements We would like to thank Henk Stoof for carefully reading the manuscript. This work was supported by the Stichting voor
Fundamenteel Onderzoek der Materie (FOM), the Netherlands
Organization for Scientific Research (NWO), and by the European
Research Council (ERC).

\bibliography{IAHEcoldatoms}

\end{document}